\documentclass[aps,prl,twocolumn,superscriptaddress,showpacs]{revtex4}
\usepackage{epsfig}
\newcommand{\beq}{\begin{equation}} 
\newcommand{\eeq}{\end{equation}} 
 
\newcommand{\ve}{\vert} 
\newcommand{\ran}{\rangle} 
\newcommand{\lan}{\langle}

\begin{document}            

\title{Frozen metastable states in ordered systems of ultrafine magnetic particles}
\author{Stefanie Russ} 
\affiliation{Institut f\"ur Theoretische Physik III, Justus-Liebig-Universit\"at
  Giessen, D-35392 Giessen, Germany}
\author{Armin Bunde} 
\affiliation{Institut f\"ur Theoretische Physik III, Justus-Liebig-Universit\"at
  Giessen, D-35392 Giessen, Germany}
\date{\today}

\begin{abstract} 
For studying the interplay of dipolar interaction and anisotropy energy in systems of ultrafine magnetic particles we consider simple cubic systems of magnetic dipoles with anisotropy axes pointing into the $z$-direction. Using Monte Carlo simulations we study the magnetic relaxation from several initial states. We show explicitely that, due to the combined influence of anisotropy energy and dipole interaction, magnetic chains are formed along the $z$-direction that organize themselves in frozen metastable domains of columnar antiferromagnetic order. We show that the domains depend explicitely on the history and relax only at extremely large time scales towards the ordered state. We consider this as an indication for the appearence of frozen metastable states also in real sytems, where the dipoles are located in a liquid-like fashion and the anisotropy axes point into random directions.
\end{abstract}
\maketitle

In the last decade, systems of ultrafine magnetic nanoparticles have received considerable interest, due both to their important technological applications (mainly in magnetic storage and recordings) and their rich and often unusual experimental behavior \cite{battle02}. An important scientific question concerns the magnetic structure of the systems \cite{binderyoung}. Several experiments on disordered magnetic materials present indications of a spin-glass phase \cite{binderyoung,morup95,jonsson95,jonsson98,rivas,kleemann1} or of a random anisotropy system \cite{Luo,taylor}, while on the theoretical side, it is a matter of controversy if, at large concentrations of nanoparticles, a spin-glass phase exists or not \cite{Andersson,ulrich,PortoPRL,porto05}. While Monte Carlo simulations on ageing \cite{Andersson} and magnetic relaxation \cite{ulrich} seem to favorize the spin-glass hypothesis, simulations of the zero-field cooling and field-cooling susceptibility showed no indication of a spin-glass phase \cite{PortoPRL}. In this paper, we use Monte-Carlo simulations (see e.g.~\cite{ewald,Nowak,Nowak2}) to study the slow magnetic relaxation from a non-equilibrium situation into the final state in ordered systems of ultrafine particles~\cite{footnote}. We find that already in this fully ordered arrangement of magnetic dipoles, the competition between anisotropy energy and dipole interaction is sufficient to produce frustration and metastable frozen states. We consider this as an indication that in the corresponding real systems of ultrafine magnetic particles, where additional frustration naturally arises as a consequence of the disorder, spinglass phases may exist.

We focus on perhaps the most basic model of magnetic nanoparticles that (i) assumes a coherent magnetization rotation within the anisotropic particles, and (ii) takes into account the magnetic dipolar interaction between them. Here, in order to get insight into the interplay of dipolar interaction and anisotropy energy, we further simplify this problem drastically by (iii) placing all magnetic particles onto the lattice points of a simple cubic lattice and (iv) orienting all anisotropy axes into the $z$-direction. It is known that the simple cubic dipolar system possesses a columnar antiferromagnetic (CAF) groundstate \cite{luttinger}, where the magnetic moments are arranged in linear chains along an arbitrary direction and each chain is surrounded by chains aligned in the opposite direction. 
In our case, the additional anisotropy energy favorizes the formation of chains along the $z$-axis, which are arranged antiferromagnetically in the $xy$-plane, similar to the case of a cubic Ising system with additional dipolar interaction \cite{kretschbind}. 

For describing the magnetic structure, we are thus led to introduce two order parameters $O_{\ell}$ and $O_{\rm{t}}$, where $O_{\ell}$ describes the magnetic order in each chain and $O_{t}$ the antiferromagnetic order between neighboring chains. Here, we study the relaxation of $O_{\ell}$ and $O_{t}$, (i) from the CAF state ($O_\ell=1$, $O_t=1$), 
and (ii) from general configurations where all dipoles are randomly oriented ($O_\ell=0$, $O_t=0$). We find that in contrast to $O_{\ell}$, the transversal order parameter $O_{t}$ depends strongly on the initial state. Only for relaxation from the CAF state, the state with lowest energy is reached quite fast, while for the general case, the system gets frozen in some intermediate disordered state. This dependence on the initial conditions becomes more pronounced in the thermodynamic limit and results, at low temperatures, in complex frozen structures that consist of several domains. In each domain, the chains are ordered in an antiferromagnetic way. 

For the numerical calculations, we use the same model as in Refs.~\cite{ulrich,PortoPRL}, where every particle $i$ of constant volume $V$ is considered to be a single magnetic domain with all its atomic magnetic moments rotating coherently. This results in a constant absolute value $\ve\mu_i\ve=M_sV$ of the total magnetic moment of each particle, where $M_s$ is the saturation magnetization. The energy of each particle consists of two contributions: anisotropy energy and dipolar interaction energy. We assume a temperature independent uniaxial anisotropy energy $E_A^{(i)}=-KV((\mu_i \vec n_i)/\ve \vec{\mu_i} \ve)^2$, where $K$ is the anisotropy constant and the unit vector $\vec n_i$ denotes the easy directions. As usual, 
the energy of the magnetic dipolar interaction between two particles $i$ and $j$ separated by $\vec r_{ij}$ is given by $E_D^{(i,j)}=(\vec \mu_i \vec \mu_j)/r_{ij}^3 -3(\vec \mu_i \vec r_{ij})(\vec \mu_j\vec r_{ij})/r_{ij}^5$. Adding up the two energy contributions and summing over all particles we obtain the total energy
\beq
E=\sum_i E_A^{(i)} +\frac{1}{2}\sum_i\sum_{j\ne i} E_D^{(i,j)}.
\eeq
In the Monte Carlo simulations we concentrate on samples of $N=L^{3}$ particles placed on a cubic lattice with periodic boundary conditions and $L$ between $4$ and $12$, where all anisotropy axes $\vec n_i$ point into the positive $z$- direction. The unitless concentration $c$ is defined as the ratio between the total volume $NV$ occupied by the particles and the volume $V_s$ of the sample. Here, we focus on the large concentration $c/c_0\approx 0.64$, where $c_0=2K/M_s^2$ is a dimensionless material-dependent constant ($c_0\sim 1.4$ for iron nitride nanoparticles \cite{ulrich,mamiya}), but we also tested samples with $c/c_0\approx 0.2$ and $0.4$ that gave qualitatively the same results. The relaxation of the individual magnetic moments per Monte Carlo step is simulated by the standard Metropolis algorithm, where the $\vec\mu_i$ are described by their two spherical coordinates $\theta_i$ and $\varphi_i$. The interaction energies are calculated using the Ewald sum method with periodic boundary conditions in $x$, $y$ and $z$-direction \cite{PortoPRL,ewald}. To study the magnetic relaxation we start from a given initial state and determine as a function of time $t$ (number of Monte Carlo steps) for each particle $i$ the angle $\theta_i$ between the magnetic moment $\vec{\mu_i}$ and the $z$-axis, from which we obtain the relevant quantities.

To quantify the relaxation process and the final state, we study the dependence of the order parameters $O_\ell(t)$ and $O_t(t)$ on $t$ (number of Monte Carlo steps). While $O_\ell \equiv \left\lan\ve m_j^{(z)}\ve\right\ran$ describes the order along the chains, $O_t\equiv -\left\lan S_j\,S_{j+\delta}\right\ran$ describes the order perpendicular to the chains. Here, $m_j^{(z)}$ is the $z$-component of the magnetic moment of the $j$th chain and $S_j=1$ if all magnetic moments of the $j$th chain point into the positive $z$-direction,
$S_j=-1$ if they point into the negative $z$-direction, and $S_j=0$ otherwise. The index $j+\delta$ denotes a nearest-neighbor chain of chain $j$ and $\lan\quad\ran$ denotes the average over all $j$, $\delta$ and over the number $N_c$ of configurations that ranges from $N_c=1000$ for the smallest to $N_c=100$ for the largest systems.  

\unitlength 1.85mm
\vspace*{0mm}
\begin{figure}
\begin{picture}(40,50)
\put(-2,25){\makebox{\includegraphics[width=4.5cm]{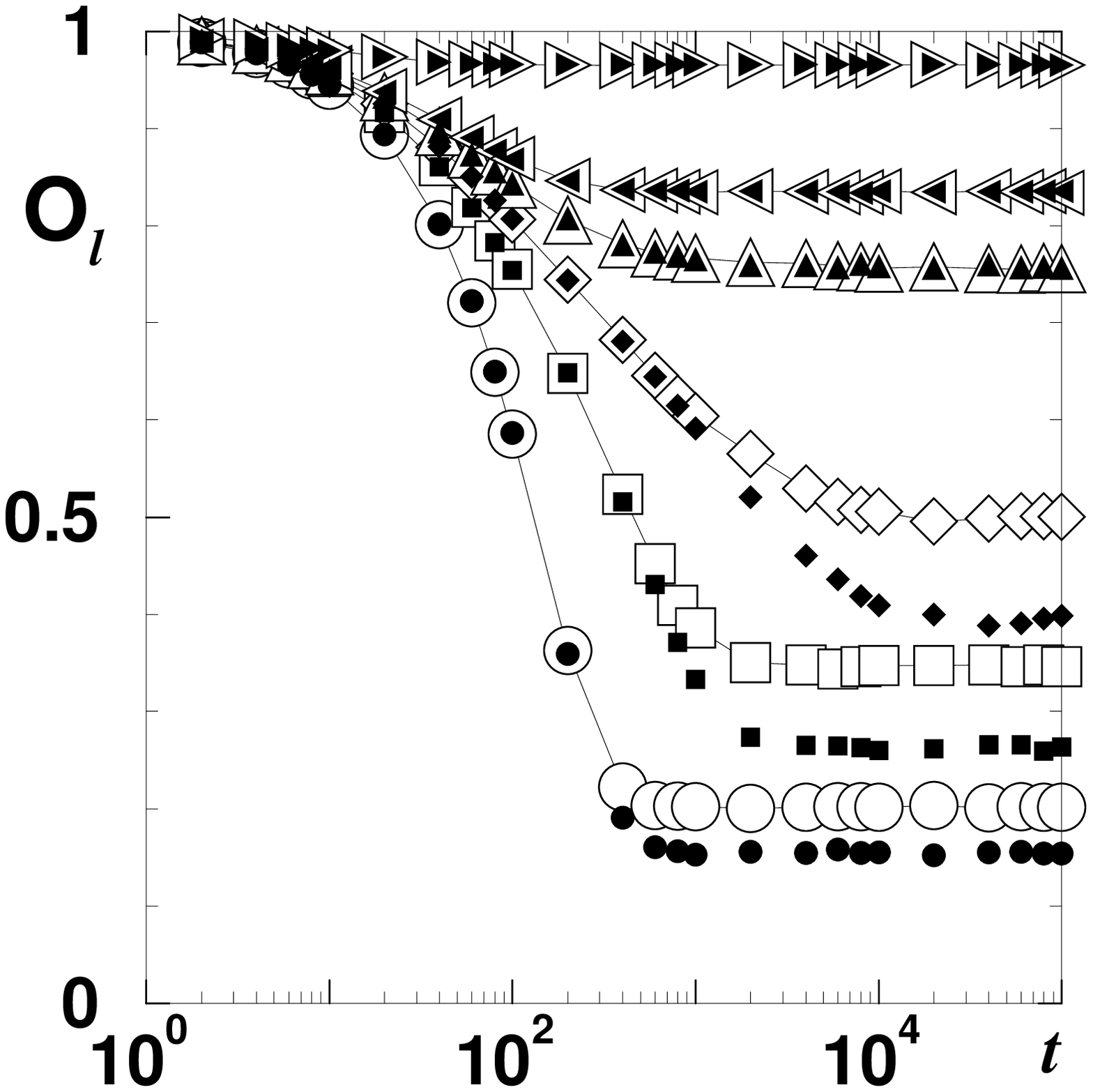}} }     
\put(22,25){\makebox{\includegraphics[width=4.5cm]{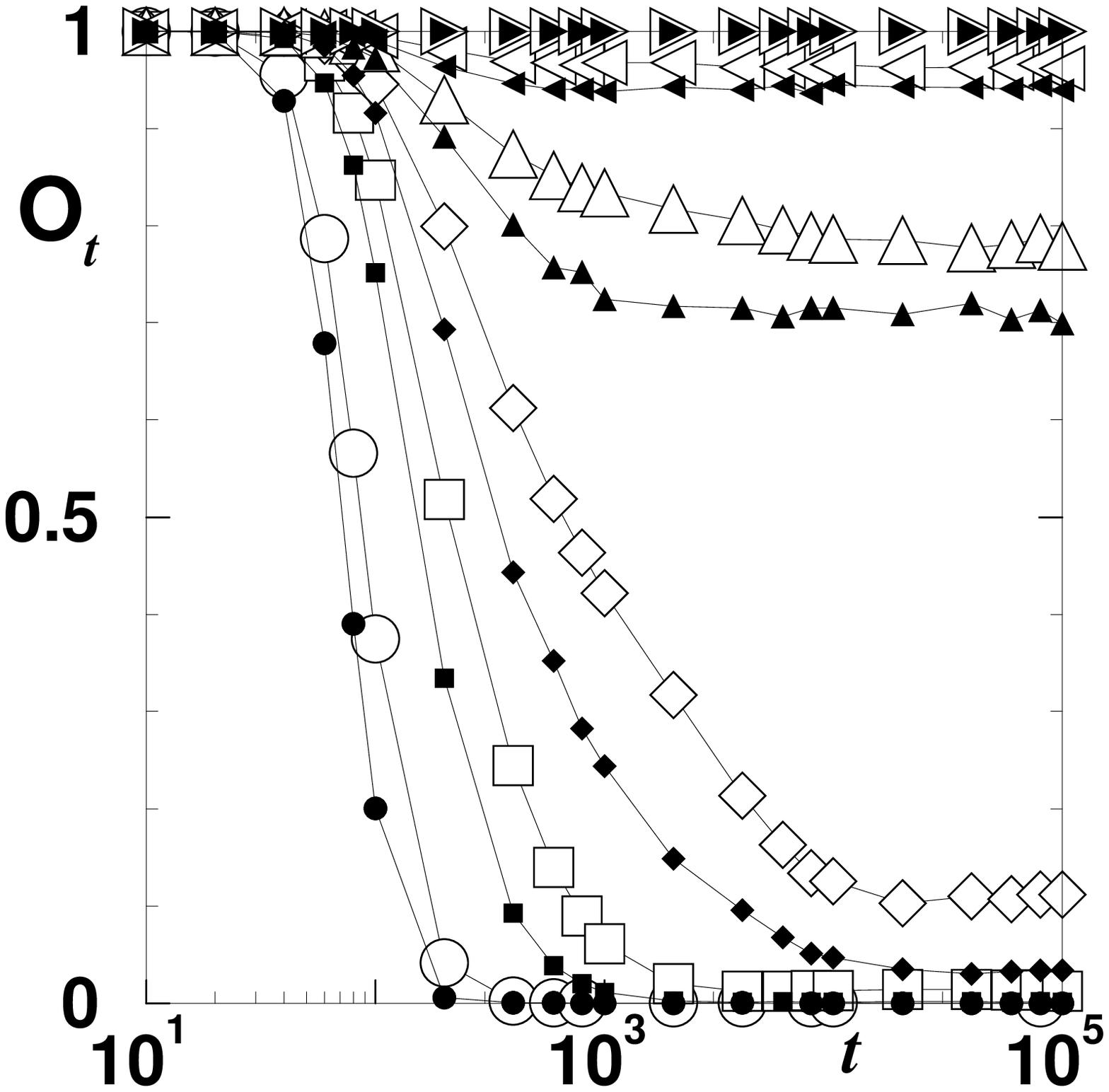}}   }   
\put(-2,0){\makebox{\includegraphics[width=4.5cm]{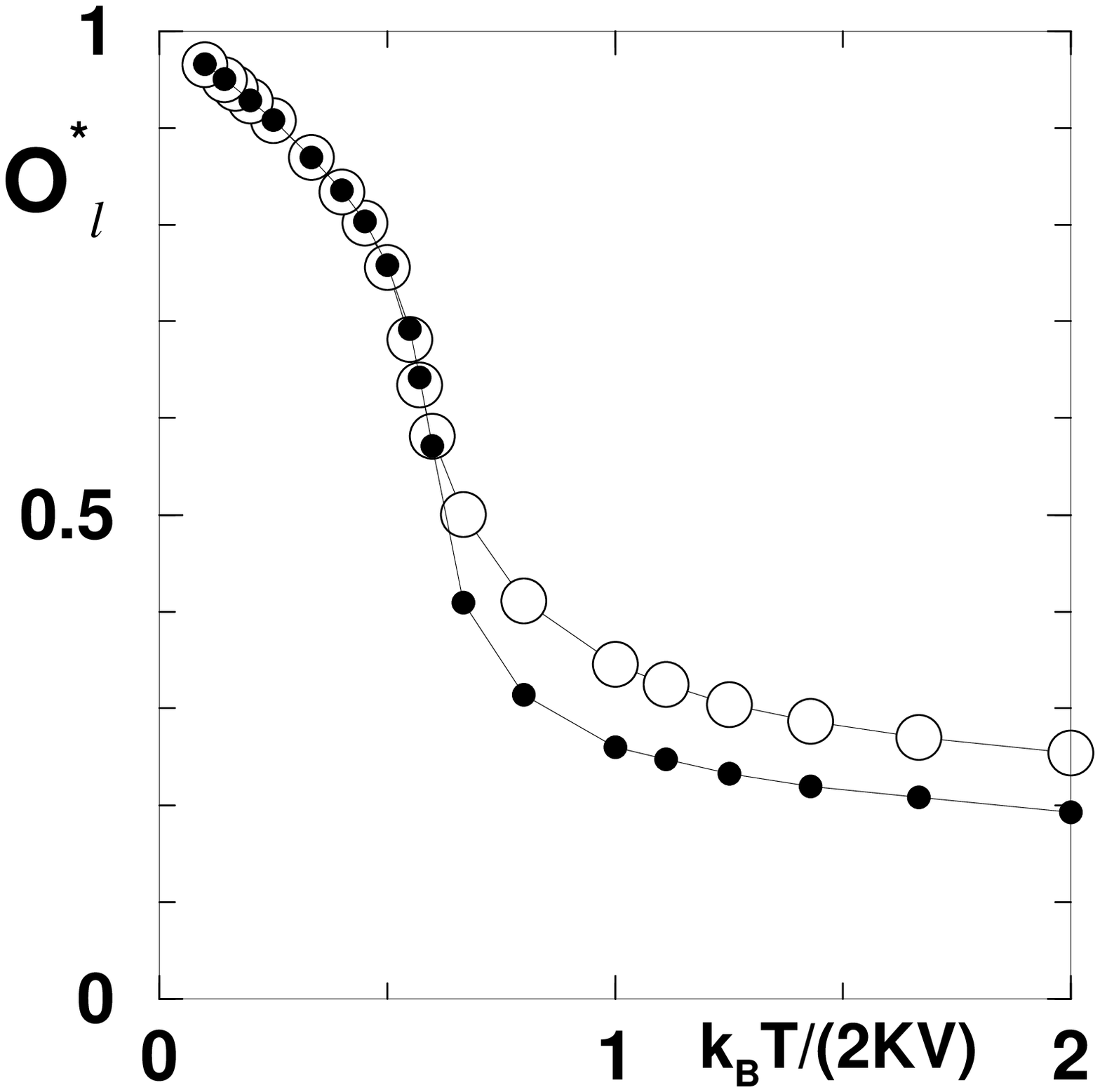}}   }   
\put(22,0){\makebox{\includegraphics[width=4.5cm]{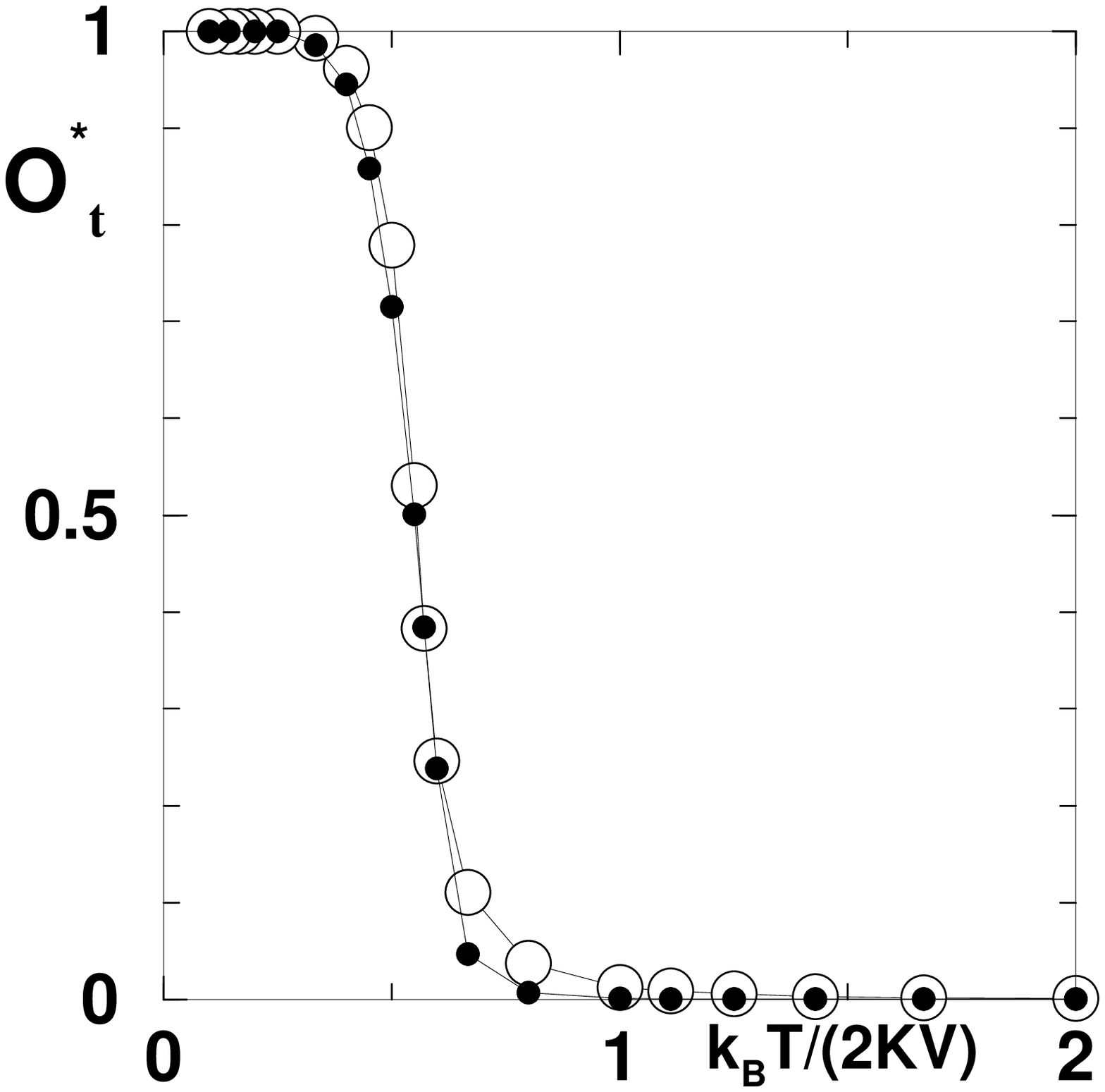}}   }   

\put(2,30){\makebox(1,1){\bf\Large (a)}} 
\put(26,30){\makebox(1,1){\bf\Large (b)}} 
\put(2,5){\makebox(1,1){\bf\Large (c)}} 
\put(26,5){\makebox(1,1){\bf\Large (d)}} 

\end{picture}
\caption[]{\small The order parameters (a) $O_{\ell}(t)$ and (b) $O_{t}(t)$ as a function of $t$ (number of Monte Carlo steps) are shown for the system sizes $L=6$ and $L=10$ when starting in the CAF state for several values of the reduced temperature $\tilde T=k_BT/(2KV)=10$ (circles), $1$ (squares), $2/3$ (diamonds), $1/2$ (triangles up), $1/2.5$ (triangles left) and $1/10$ (triangle right), where $T$ is the temperature, $k_B$ the Boltzmann constant, $K$ the anisotropy constant and $V$ the particle volume. In (c,d) $O_\ell^*$ and $O_t^*$ are shown for fixed $t=10^4$ as a function of $\tilde T$. }
\label{bi:af}
\end{figure}

In order to identify a temperature above which the ferromagnetic order inside the chains and the antiferromagnetic order between the chains cannot be preserved, we first consider relaxation from the CAF state. 
Figures~\ref{bi:af}(a,b) show $O_{\rm{\ell}}(t)$ and $O_{\rm{t}}(t)$ for systems of $N=6^{3}$ (white symbols) and $10^{3}$ (black symbols) particles as a function of $t$, for several values of the reduced temperature $\tilde T=k_BT/(2KV)$, where $2KV$ is the height of the anisotropy barrier. One can see that $O_\ell(t)$ and $O_t(t)$ reach final plateau values $O_\ell^*$ and $O_t^*$ quite fast. Figures~\ref{bi:af}(c,d) show these values as a function of $\tilde T$. The figures show that for small temperatures, $O_\ell^*$ and $O_t^*$ are close to $1$ and decay rapidly to much smaller values in a narrow temperature regime around $\tilde T\approx 0.5$. In both cases, the width of the transition regime shrinks with increasing system size. From the inflection point, we can identify the critical temperature $T_c\approx 0.5$ above which the longitudinal and transversal order breaks down. The figures show that both, the order along the chains and the order in the $xy$-plane disappears at the same transition point. We can also see in the figure that only as long as $T$ is below $T_c$, the order parameters do not depend on the system size $L$. For $T>T_c$, $O_t$ decreases with $L$, which reflects the fact that the chains have started to break. 

\unitlength 1.85mm
\vspace*{0mm}
\begin{figure}
\begin{picture}(40,23)
\put(-2,0){\makebox{\includegraphics[width=4.5cm]{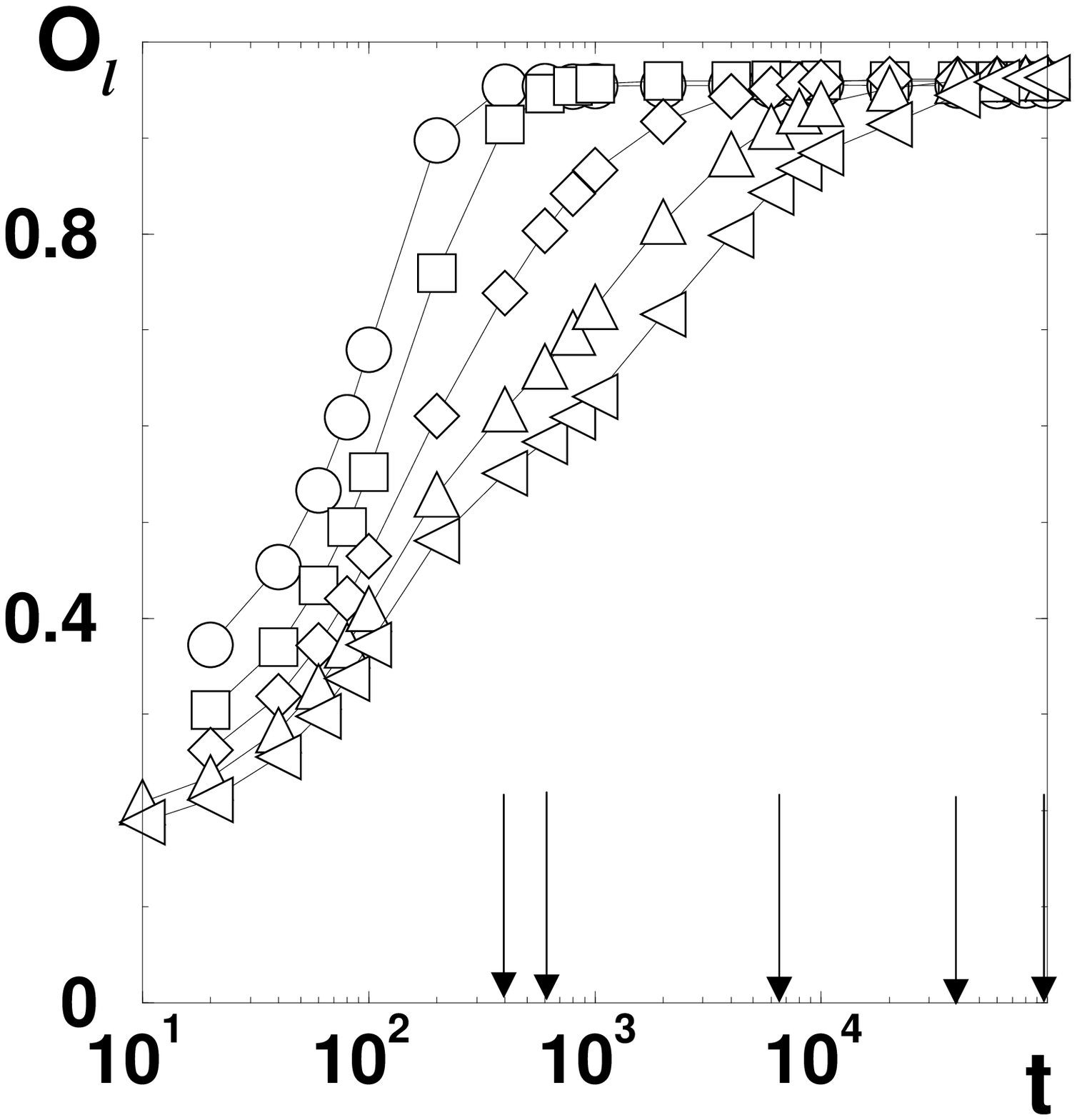}}   }   
\put(22,0){\makebox{\includegraphics[width=4.5cm]{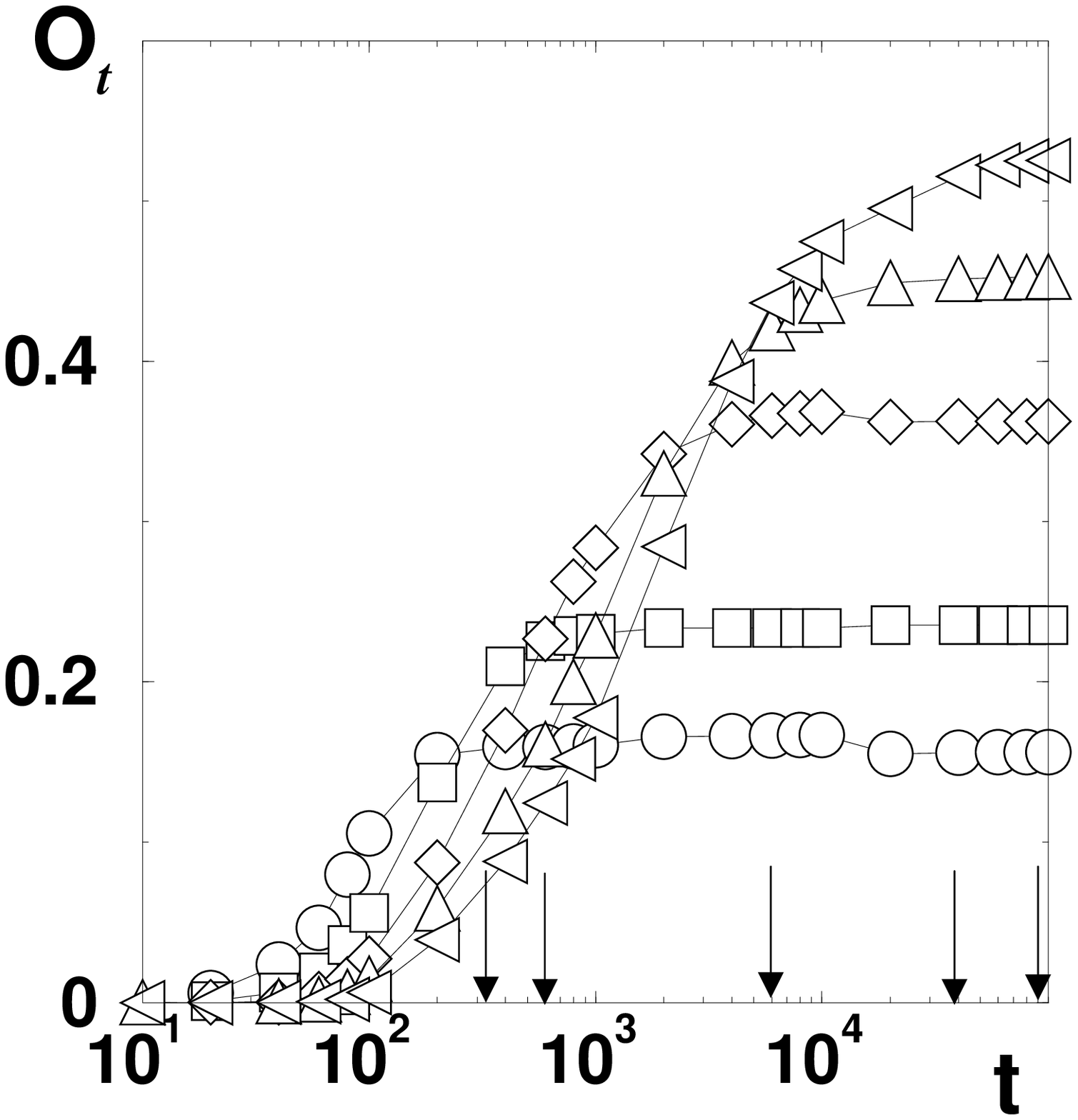}}   }   

\put(1,20){\makebox(1,1){\bf\Large (a)}} 
\put(26,20){\makebox(1,1){\bf\Large (b)}}

\end{picture}
\caption[]{\small The order parameters (a) $O_\ell(t)$ and (b) $O_t(t)$ are plotted versus the time $t$ (number of Monte Carlo steps) when starting in random configurations for the temperature $\tilde T=1/10$ and for $L=4$ (circles), $6$ (squares), $8$ (diamonds), $10$ (triangles up) and $L=12$ (triangles left). The arrows indicate the approximate crossover times $t_1$ for the system sizes (from left to right) $L=4$ to $L=12$.
}
\label{bi:sizes1}
\end{figure}

\unitlength 1.85mm
\vspace*{0mm}
\begin{figure}
\begin{picture}(40,30)
\put(0,0){\makebox{\includegraphics[width=6cm]{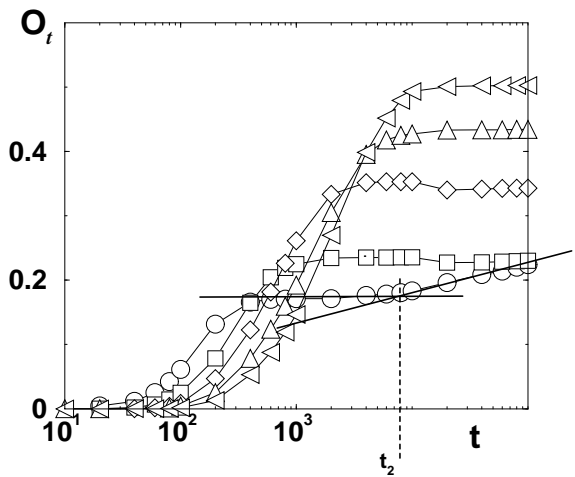}}   }   
\end{picture}
\caption[]{\small The order parameters $O_t(t)$ are plotted versus the time $t$ (number of Monte Carlo steps) when starting in random configurations for several $L$ from $L=4$ to $L=12$ (same symbols as in Fig.~\ref{bi:sizes1}) and for the temperature $\tilde T=1/5$. The second crossover time $t_2$ is shown for $L=4$. 
}
\label{bi:sizes2}
\end{figure}

Next, we consider the relaxation from a random initial configuration. We start with $\tilde T=1/10$, a temperature well below $\tilde T_c$. Figure~\ref{bi:sizes1} shows that again both, $O_\ell(t)$ and $O_t(t)$ behave in a similar way. They first increase with $t$ and then reach a plateau value at the same crossover time $t_1(L)$ that increases monotonically with $L$ (see the arrows in Fig.~\ref{bi:sizes1}). But while $O_\ell$ reaches an $L$-independent plateau value (which agrees with the equilibrium value $O_\ell^*$ for $\tilde T=1/10$ of Figs.~\ref{bi:af}(a,c)), the plateau values of $O_t$ continue to increase with the system size and seem to approach the corresponding value $O_t^*$ from Figs.~\ref{bi:af}(b,d) only in the thermodynamic limit. 

\unitlength 1.85mm
\begin{figure}
\begin{picture}(40,75)

\put(-2,12){\makebox{\includegraphics[width=1.7cm]{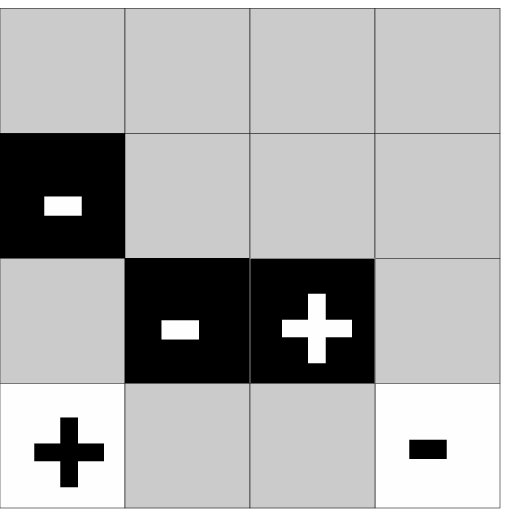}}   }   
\put(10,12){\makebox{\includegraphics[width=1.7cm]{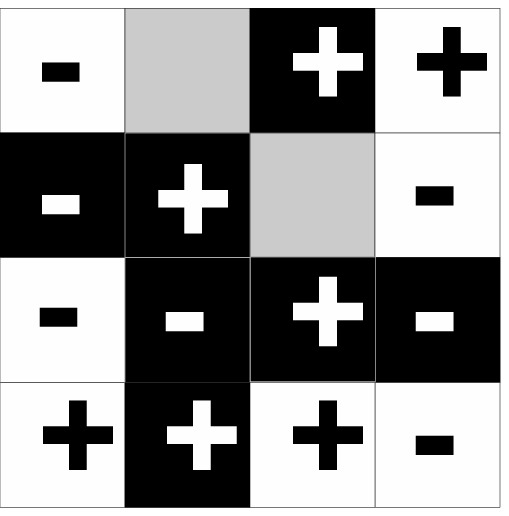}}   }   
\put(-2,0){\makebox{\includegraphics[width=1.7cm]{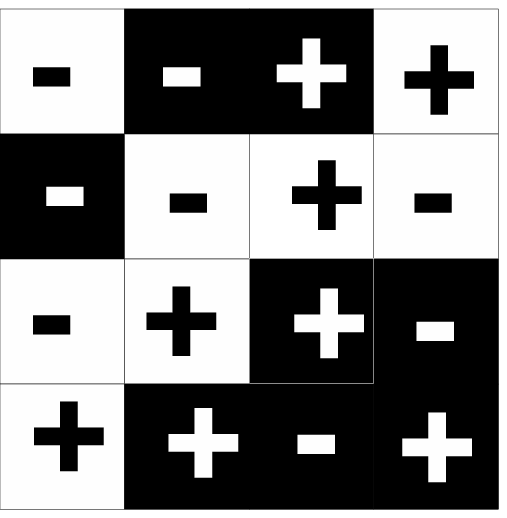}}   }   
\put(10,0){\makebox{\includegraphics[width=1.7cm]{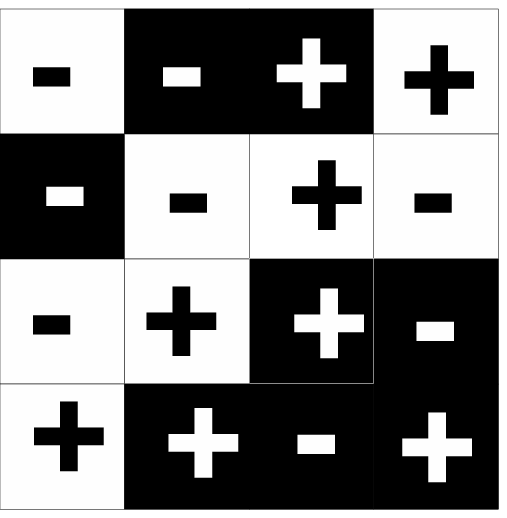}}   }   

\put(-2,51.5){\makebox{\includegraphics[width=4cm]{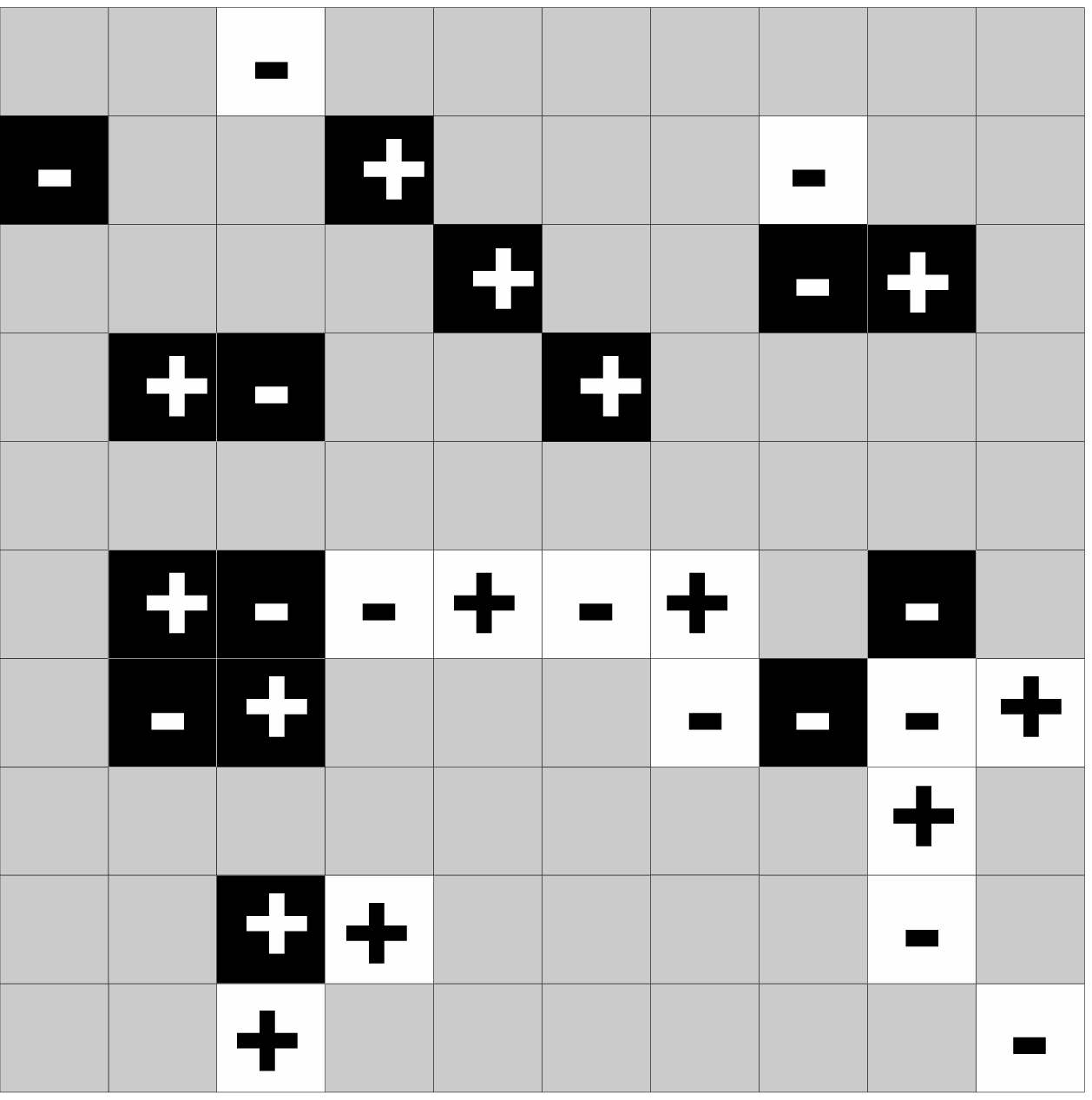}}   }   
\put(21,51.5){\makebox{\includegraphics[width=4cm]{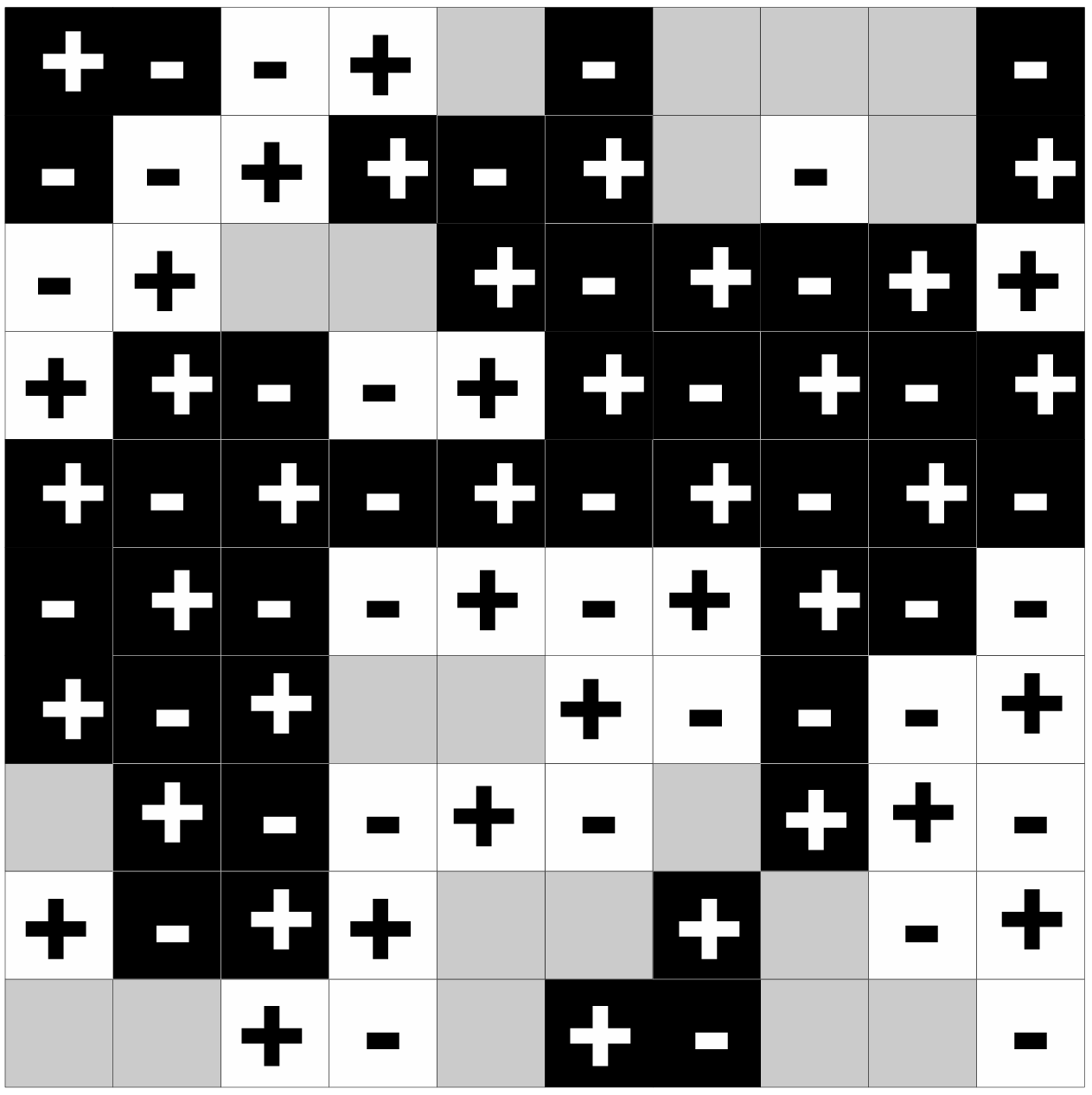}}   }   
\put(-2,26){\makebox{\includegraphics[width=4cm]{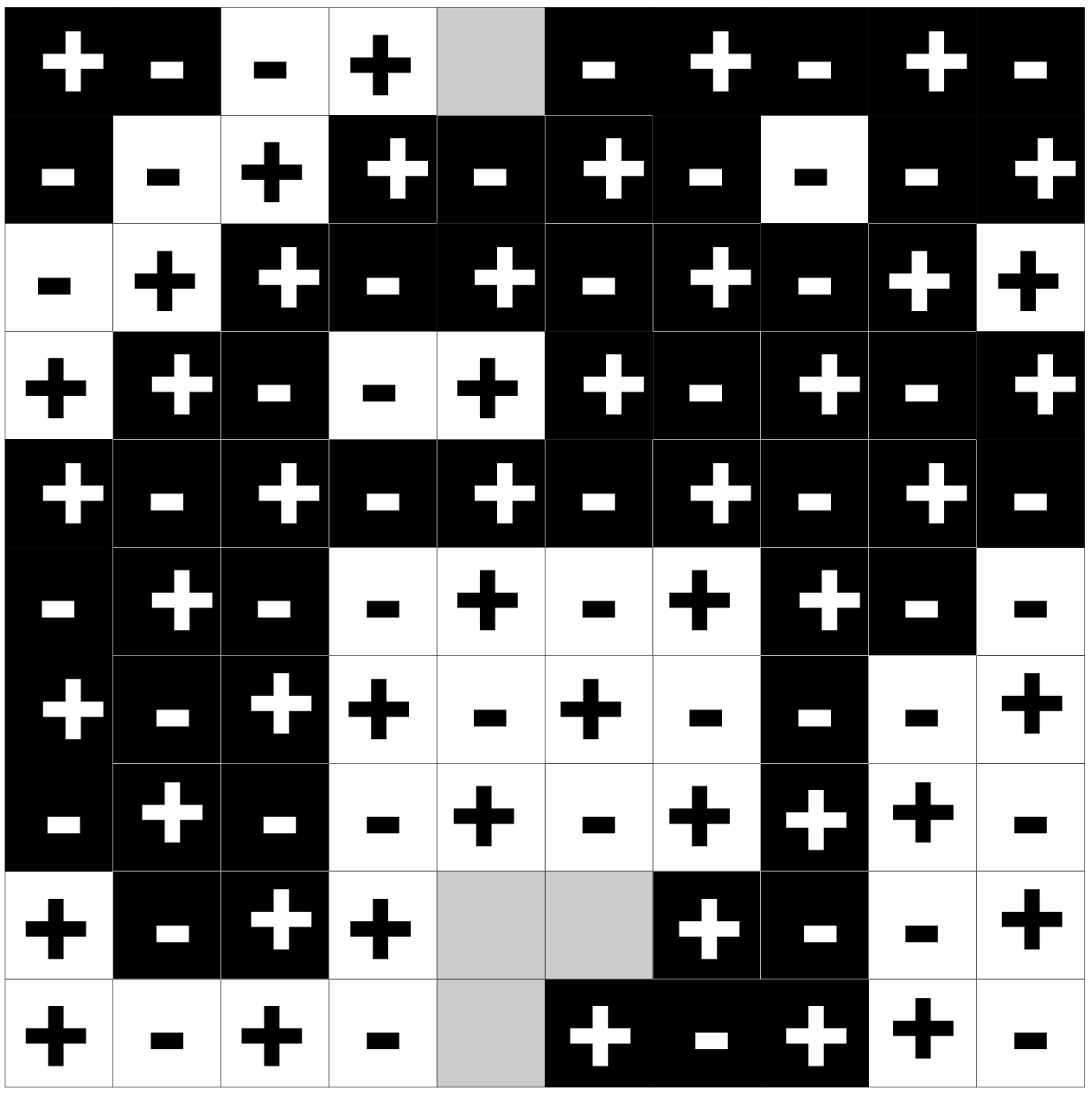}}   }   
\put(21,26){\makebox{\includegraphics[width=4cm]{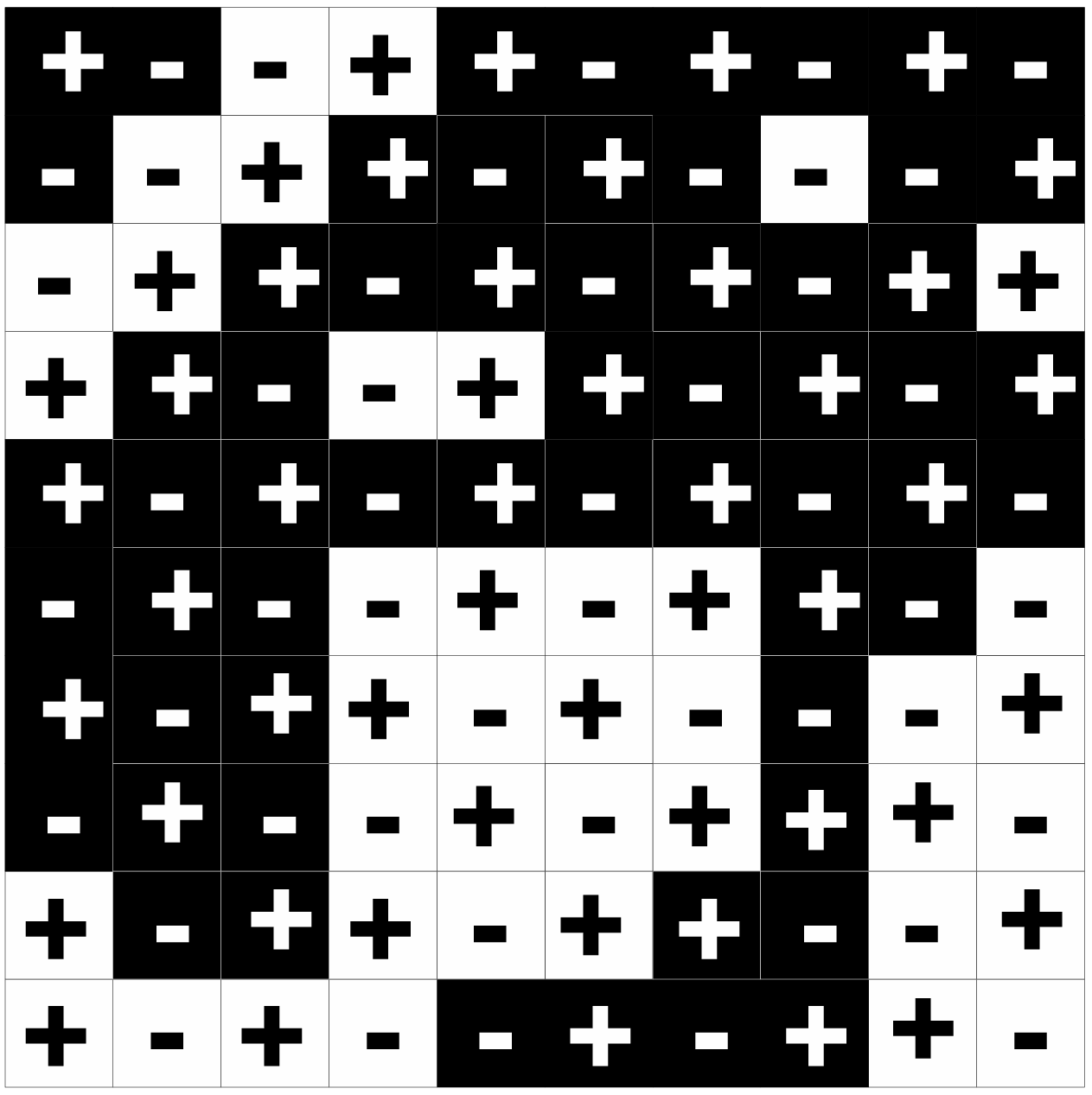}}   }   
\put(23,-2){\makebox{\includegraphics[width=4cm]{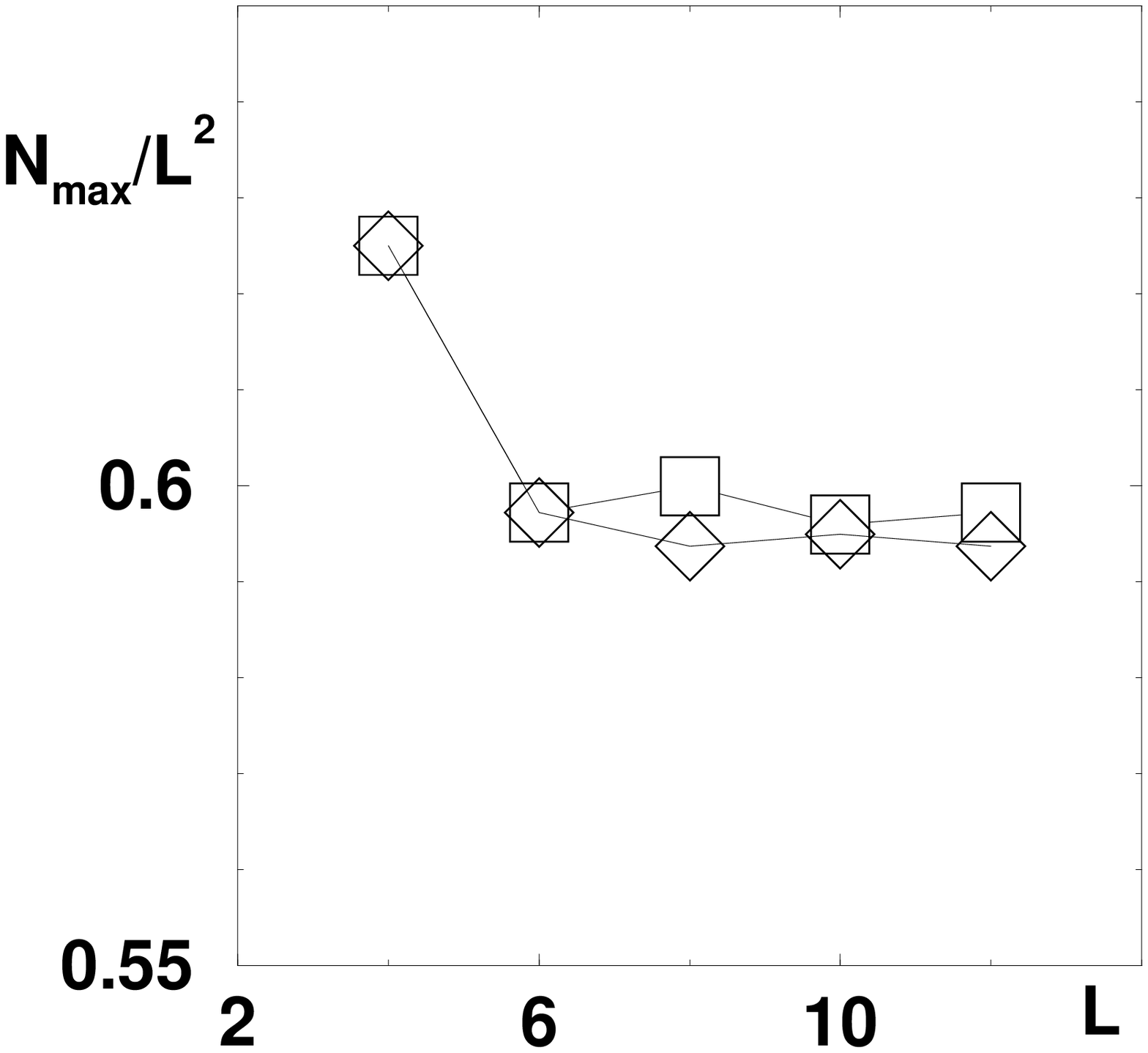}}   }   

\put(8,75){\makebox(1,1){\bf\Large (a)}} 
\put(30,75){\makebox(1,1){\bf\Large (b)}} 
\put(8,49.){\makebox(1,1){\bf\Large (c)}} 
\put(30,49.){\makebox(1,1){\bf\Large (d)}} 
\put(2,23.){\makebox(1,1){\bf (e)}} 
\put(14,23.){\makebox(1,1){\bf (f)}} 
\put(2,10.5){\makebox(1,1){\bf (g)}} 
\put(14,10.5){\makebox(1,1){\bf (h)}} 
\put(30,22.){\makebox(1,1){\bf\Large (i)}} 

\end{picture}
\caption[]{\small Visualization of the antiferromagnetic order in the $xy$-plane for systems at $\beta=10$ when starting in a random configuration.
(a-d) One $L=10$ system at fixed $t$ (number of Monte Carlo steps), $t=10^2$, $10^3$, $10^4$ and  $t=10^5$, (e-h) one $L=4$ system after $t=20, 10^2, 2 \cdot 10^3$ and $10^5$. The complete chains are indicated by $+$ or $-$ signs, depending on the direction of the chain. Antiferromagnetic domains are shown in black and white. Domains of the same color fit to each other and are allowed to emerge, whereas domain walls exist between clusters of different colors. Sites, where chains have not yet been built are indicated by the grey color. (e) The average relative size $N_{\rm{max}}/L^{2}$ of the largest cluster at $t$ with $t_1<t<t_2$ is plotted versus the size $L$ for $\tilde T=1/10$ (squares) and $\tilde T=1/20$ (diamonds).
}
\label{bi:pattern}
\end{figure}

Therefore, we arrive at the following picture for $T<T_c$: Below $t_1$, the system is disordered, while above $t_1$, all magnetic moments are ordered in chains along the $z$-direction. But in contrast to the relaxation from the CAF state (Fig.~\ref{bi:af}), the chains are not (yet) ordered in an antiferromagnetic way. Accordingly, the system has not yet reached its (equilibrium) state of minimum energy and must be considered as metastable. We therefore expect that there exists a second crossover time $t_2$, which can be roughly identified as the inverse of the probability $P(L,T)$ that a chain flips into a more favorable direction. Accordingly, above $t_2\sim P^{-1}(L,T)$ chain flips will dominate the relaxation and this leads to a further increase of $O_t(t)$ towards its equilibrium value $O_t^*$ (see Fig.~\ref{bi:af}(d)). Since $P(L,T)$ decays drastically with decreasing $T$ and increasing $L$, one has to go to small system sizes and comparatively large temperatures (below $T_c$) in order to find $t_2$. Indeed, Fig.~\ref{bi:sizes2} shows that at the intermediate temperature $\tilde T=1/5$, $t_2$ can be observed only for the smallest system size $L=4$. We like to note that the metastable state shown here differs from the typical metastable spin-glass structure identified in \cite{kretschbind} for an Ising system, where the up- and down-columns are arranged in a completely random fashion and $O_t$ should be equal to zero. Instead, we will show in the following that in our system, different domains grow from different seeds.

We now consider the $L$-dependence of $O_t(t)$. The increase of the plateau values with $L$ in Figs.~\ref{bi:sizes1}(b) and \ref{bi:sizes2} seems to suggest that in the thermodynamic limit, the system reaches the CAF groundstate already at $t_1(L)$. To see that this is not the case, we illustrate the relaxation process in Fig.~\ref{bi:pattern} for $\beta=10$ and two different system sizes $L=4$ and $L=10$. To visualize the antiferromagnetic order, we follow the definition of $O_t$: each of the $L^{2}$ sites in the $xy$-plane can be either a ''+'' site or a ''-'' site, if all magnetic moments in the chain point into the positive or negative $z$-direction, respectively, or a ''0'' site if this is not yet the case (grey sites), in close analogy to the definition of $S_j$. The figure shows that in the initial time steps, for both sizes, only distant isolated chains have been formed. Due to the dipolar interaction, chains are more likely formed in the neighborhood of another chain. This way, small (columnar) antiferromagnetic domains develop that grow further with increasing time. When different domains contact each other, they either emerge and form larger domains (if they fit to each other) or establish quite stable domain walls between them. In Fig.~\ref{bi:pattern}, both types of domains are in black and white, respectively, such that domains of the same color fit to each other and are in principle allowed to emerge. The figure illustrates the change of the domains (i) with $t$ (number of Monte Carlo steps) and (ii) with system size $L$.

First, Fig.~\ref{bi:pattern} shows that below $t_1$, the longitudinal order is not yet fully established and in rare cases (particularly for small system sizes) complete chains can still flip into the other direction (compare Figs.~\ref{bi:pattern}(f) and (g)). Above $t_1$, chain flips are too rare to be observed and the structure is practically frozen-in, until at considerably larger time scales (above $t_2$, not shown here) chain flips may again occur. However, since $t_2$ increases drastically with $L$, the structure is practically frozen-in above $t_1$ for reasonably large system sizes.  The reason for this glassy-like phenomenon is the competition between the longitudinal order along the chains and the transversal order in the plane that leads to frustration of the single magnetic moments inside the chains.

Second, also the monotonous increase of the plateau value of $O_t$ with $L$ in Figs.~\ref{bi:sizes1}(b) and \ref{bi:sizes2} can be understood qualitatively from Fig.~\ref{bi:pattern}. According to the figure, the system consists of large clusters and of small inclusions inside them. Above $t_1$, the fraction of the domain wall sites decreases with increasing $L$, and thus $O_t$ is enhanced. (Compare Figs.~\ref{bi:pattern}(h) and \ref{bi:pattern}(d): in the first case, all sites except one belong to a domain wall, whereas in the case of $L=10$ many interior sites exist.) Nevertheless, the system does not reach an homogeneous antiferromagnetic order. Instead, as it is shown in Fig.~\ref{bi:pattern}(i), we have found numerically that at small temperatures, the fraction of sites occupied by the largest cluster, $S_{\rm{max}}/L^{2}$, has a size-independent value that is close to $0.6$ for $\tilde T=1/10$ and $1/20$. This means that the relative size of the largest cluster does not grow with $L$ on dispense of the smaller clusters, i.e., no unique domain is reached. 

From the preceeding it is evident that different initial conditions will lead to different domain structures in the plateau regime. When we start e.g. with a random chain-like structure consisting of certain (columnar) antiferromagnetic domains, the first crossover time $t_1$ vanishes as for the relaxation from the CAF and the initial state will freeze in. Therefore, in the plateau regime, the structure of the frozen configuration depends strongly on the history.

In summary, we have shown that already an ordered system of ultrafine magnetic particles where each particle is located at a lattice site and its anisotropy axes are oriented parallel to each other, shows complex dynamical behavior with the formation of frozen history-dependent states, where linear magnetic chains are formed that are quite stable and act as seeds for the formation of the frozen domain structure. Accordingly, the complex behavior results from the interplay between dipole interaction and anisotropy energy. We expect that also in real sytems of ultrafine magnetic particles, where the dipoles are located in a liquid-like fashion and the anisotropy axes point into random directions, similar linear structures are formed in the beginning of the process and act as seeds for the formation of larger frozen-in structures in the same way as the chains in the ordered systems. We thus take the results for the ordered magnetic structure as an indication that also in the corresponding disordered system, where frustration arises in a quite natural way, spinglass phases exist at low temperatures.

We gratefully acknowledge financial support from the Deutsche 
Forschungsgemeinschaft


\begin{thebibliography} {50}
\bibitem{battle02} X. Batlle and A. Labarta, J. Phys. D 35, R15 (2002). 
\bibitem{binderyoung} K. Binder and A. P. Young, Rev. Mod. Phys. {\bf 58}, 801 (1986).
\bibitem{morup95} S. Morup, F. Bodker, P.~V.~Hendriksen, and S. Linderoth, Phys. Rev. B {\bf 52}, 287 (1995).
\bibitem{jonsson95} T.~Jonsson, P.~Svedlindh, and M.~Hansen, Phys. Rev. Lett. {\bf 81}, 3976 (1998).
\bibitem{jonsson98} T. Jonsson, J. Mattsson, C. Djurberg, F. A. Khan, P. Nordblad, and  Phys. Rev. Lett. {\bf 75}, 4138 (1995).
\bibitem{rivas} F.~Rivadulla, M.~A.L\'opez-Quintela, and J.~Rivas, Phys. Rev. Lett. {\bf 93}, 167206 (2004).
\bibitem{kleemann1} X. Chen, S. Sahoo, W. Kleemann, S. Cardoso and P. P. Freitas, Phys. Rev. B {\bf 70} 172411 (2004).
\bibitem{Luo} W. Luo, S. R. Nagel, T. F. Rosenbaum, and R. E. Rosensweig, Phys. Rev. Lett. {\bf 67}, 2721 (1991). 
\bibitem{taylor} D.~R.~Taylor and W.~J.~L.~Buyers, Phys. Rev. B {\bf 54}, R3734 (1996).
\bibitem{Andersson} J.-O. Andersson. C. Djurberg, T. Jonsson, P. Svedlindh, and P. Nordbladet, Phys. Rev. B {\bf 56}, 13983 (1997).
\bibitem{ulrich} M. Ulrich, J. Garc´{i}a-Otero, J. Rivas, and A. Bunde, Phys. Rev. B {\bf 67}, 024416 (2003).
\bibitem{PortoPRL} J. Garc´{i}a-Otero, M. Porto, J. Rivas, and A. Bunde, Phys. Rev. Lett. {\bf 84}, 167 (2000).
\bibitem{porto05} M. Porto, Eur. Phys. J. B {\bf 45}, 369 (2005).
\bibitem{ewald} M. P. Allen and D. J. Tildesley, {\it Computer Simulation of Liquids} (Clarendon, Oxford, 1987).
\bibitem{Nowak} U. Nowak, R. W. Chantrell, and E. C. Kennedy, Phys. Rev. Lett. {\bf 84}, 163 (2000).
\bibitem{Nowak2} D.~Hinzke and U. Nowak, Phys. Rev. B {\bf 61}, 6734 (2000).
\bibitem{footnote} As has been shown in Ref.~\cite{Nowak} for systems of non-interacting and in \cite{Nowak2} of interacting magnetic particles, the Monte Carlo method is well suited to treat the relaxation of magnetic particles when the Monte Carlo time steps are large compared with the precession time of the magnetic moments and a mapping to a real time scale is possible. Recently, a Monte Carlo method that also includes the precessional motion has been introduced in \cite{cheng06}. Alternative molecular-dynamic simulations are not useful for treating slow relaxation processes, since the accessible time scales are too short.
\bibitem{luttinger} J. M. Luttinger and L. Tisza, Phys. Rev. {\bf 70}, 954 (1946).
\bibitem{kretschbind} R. Kretschmer and K. Binder, Z. Phys. B {\bf 34}, 375 (1979).
\bibitem{mamiya} H.~Mamiya, I.~ Nakatani and T.~Furubayashi; Phys. Rev. Lett. {\bf 80} 177 (1998).
\bibitem{cheng06} X.~Z.~Cheng, M.~B.~A.~Jail, H.~K.~Lee and Y.~Okabe, Phys. Rev. Lett. {\bf 96}, 067208 (2006).
\end{thebibliography}
\end{document}